\documentclass{aa}  

\usepackage{natbib}
\bibpunct{(}{)}{;}{a}{}{,} 
\usepackage{graphicx}
\usepackage{txfonts}
\usepackage{txfonts}
\usepackage{color}
\usepackage{colortbl}
\usepackage{placeins}
\usepackage{float}
\usepackage{subfigure}
\usepackage{comment}
\begin{document} 

   \title{Studies of stationary features in jets: BL Lacertae II. Trajectory reversals and superluminal speeds on sub-parsec scales}

   \authorrunning{Arshakian et al.}
   \titlerunning{Trajectory reversals and superluminal speeds on sub-parsec scales}
   \subtitle{}

   \author{T.G. Arshakian
          \inst{1,2,3},
          L.A. Hambardzumyan
          \inst{2,3}, 
          A.B. Pushkarev
          \inst{4,5},
          \and 
          D.C. Homan
          \inst{6}
          }

   \institute{I. Physikalisches Institut, Universität zu Köln, Zülpicher Strasse 77, Köln, Germany, \\
              \email{arshakian@ph1.uni-koeln.de}
         \and
            Byurakan Astrophysical Observatory after V.A. Ambartsumian, Aragatsotn Province 378433, Armenia,
        \and
            Astrophysical Research Laboratory of Physics Institute, Yerevan State University, 1 Alek Manukyan St., Yerevan, Armenia
        \and
            Crimean Astrophysical Observatory, 298409 Nauchny, Crimea, Russia
        \and 
            Institute for Nuclear Research of the Russian Academy of Sciences, 60th October Anniversary Prospect 7a, Moscow 117312, Russia
        \and
            Department of Physics and Astronomy, Denison University, Granville, OH 43023, USA
             }

   \date{Received xxx; accepted xxx}


  \abstract  
  {High-resolution VLBI observations revealed a quasi-stationary component (QSC) in the relativistic jets of many blazars, which represents a standing recollimation shock. The VLBA monitoring of the BL Lacertae jet at 15~GHz shows the QSC at a projected distance of about 0.26~mas from the radio core.} 
   {We study the trajectory and kinematics of the QSC in BL Lacertae on sub-parsec scales using 15~GHz VLBA data of 164 observations over 20 years from the MOJAVE program and 2~cm VLBA Survey.}
   {To reconstruct the QSC's intrinsic trajectory, we use moving average and trajectory refinement procedures to smooth out the effects of core displacement and account for QSC positioning errors.}
  {We identified 22 QSC reversal patterns with a frequency of $\sim 1.5$ per year. Most reversals have an acute angle $<90\degr$ and few have a loop-shaped or arc-shaped trajectory. Where observed, combinations of reversals show reversible and quasi-oscillatory motion. We propose a model in which a relativistic transverse wave passes through the QSC, generating a short-lived reverse motion, similar to the transverse motion of a seagull on a wave. According to the model, relativistic waves are generated upstream and the reverse motion of the QSC is governed by the amplitude, velocity and tilt of the wave as it passes through. The apparent superluminal speeds of the QSC ($\sim 2\,c$) are then due to the relativistic speed of the jet's transverse wave ($<0.3\,c$ in the host galaxy rest frame) combined with the relativistic motion towards the observer. The measured superluminal speeds of QSC
   indirectly indicate the presence of relativistic transverse waves, and the size of the QSC scattering on the sky is proportional to the maximum amplitude of the wave. We find that most of the transverse waves are twisted in space. In the active state of the jet, the directions of the twisting waves are random, similar to the behaviour of the wave in a high-pressure hose, while in the jet stable state, the wave makes quasi-oscillations with regular twisting. }
    {The study of QSC dynamics in BL Lac-type blazars is important for evaluating the physical characteristics of relativistic transverse jet waves. The latter have important implications for jet physics and open up possibilities for modelling the physical conditions and location in the jet necessary for the excitation of relativistic transverse waves. 
    }


   \maketitle
%

\section{Introduction}
Active galactic nuclei of blazars (flat-spectrum BL Lac objects, quasars, and radio galaxies) are the most energetic sources of radiation generated by accretion disc and relativistic jets of plasma material. High-resolution VLBA (Very Long Baseline Array) monitoring of jets at 15~GHz and 43~GHz shows the existence of quasi-stationary radio components (QSC) in many blazars \citep[see][and references therein]{cohen14}. Analysis of the data of the 110 VLBA 15~GHz observations of the BL Lac object, \cite{cohen14} revealed a brightest quasi-stationary radio component (labeled as C7) next to the core at a distance of 0.26~mas and moving superluminal components, which emanate after C7 component. \cite{cohen15} studied the kinematics of the jet ridge lines downstream of C7 and reported transverse waves propagating downstream at relativistic speeds. Based on the similarity of the change in position angle of C7 and the jet ridge line at $\sim 1$~mas, they proposed a model of a rapidly shaking whip at C7, which represents a recollimation shock and acts as a jet nozzle, trapping the jet stream and exciting superluminal transverse waves propagating downstream. 
\cite{arshakian20} reported that during the active state of the jet, large displacements of C7 are accompanied by the generation of strong transverse waves at C7, while quasi-sinusoidal waves with amplitude $\lessapprox 0.02$~mas are generated when the jet is in a steady state. They showed that the apparent motion of C7 is a combination of the core displacements and the intrinsic motion of C7, and that the contribution of both factors is equally important. They concluded that to study the intrinsic trajectory and velocity of C7, the core displacements and asymmetric positioning errors of C7 must be properly accounted for. The speeds of C7 were found to be superluminal ($>1.2\,c$), which was unexpected since the motion of the C7 occurs in a plane almost perpendicular to the line of sight and relativistic effects should be relaxed.

To address above issues, we conduct a thorough study of the trajectory and kinematics of the quasi-stationary component in BL Lac using 20 years of 15~GHz VLBA monitoring data from the MOJAVE program \citep[Monitoring Of Jets in Active Galactic nuclei with VLBA Experiments,][]{lister09}. Section~\ref{sec:observational data} describes briefly observational data. In Section~\ref{sec:C7 trajectory analysis} we analyse the C7 trajectory and classify the C7 reverse trajectories, and present a model for reversible motion in Section~\ref{sec:C7 model}. We describe the C7 superluminal speeds in the framework of proposed C7 model in Section~\ref{sec:C7 superluminal speeds} and discuss the model in Section~\ref{sec:discussions}. 

For the BL Lac at redshift $z = 0.0686$ \citep{vermeulen95}, the linear scale is 1.296~pc~mas$^{-1}$, assuming a flat cosmology with $\Omega_{m} = 0.27$, $\Omega_{\Lambda} = 0.73$, and $H_0 = 71$~km~ s$^{-1}$~Mpc$^{-1}$ \citep{komatsu09}.

\section{Observational data and errors of measurements}
\label{sec:observational data}
\begin{figure}[hpt]
\centering
\includegraphics[width=9cm,angle=0] {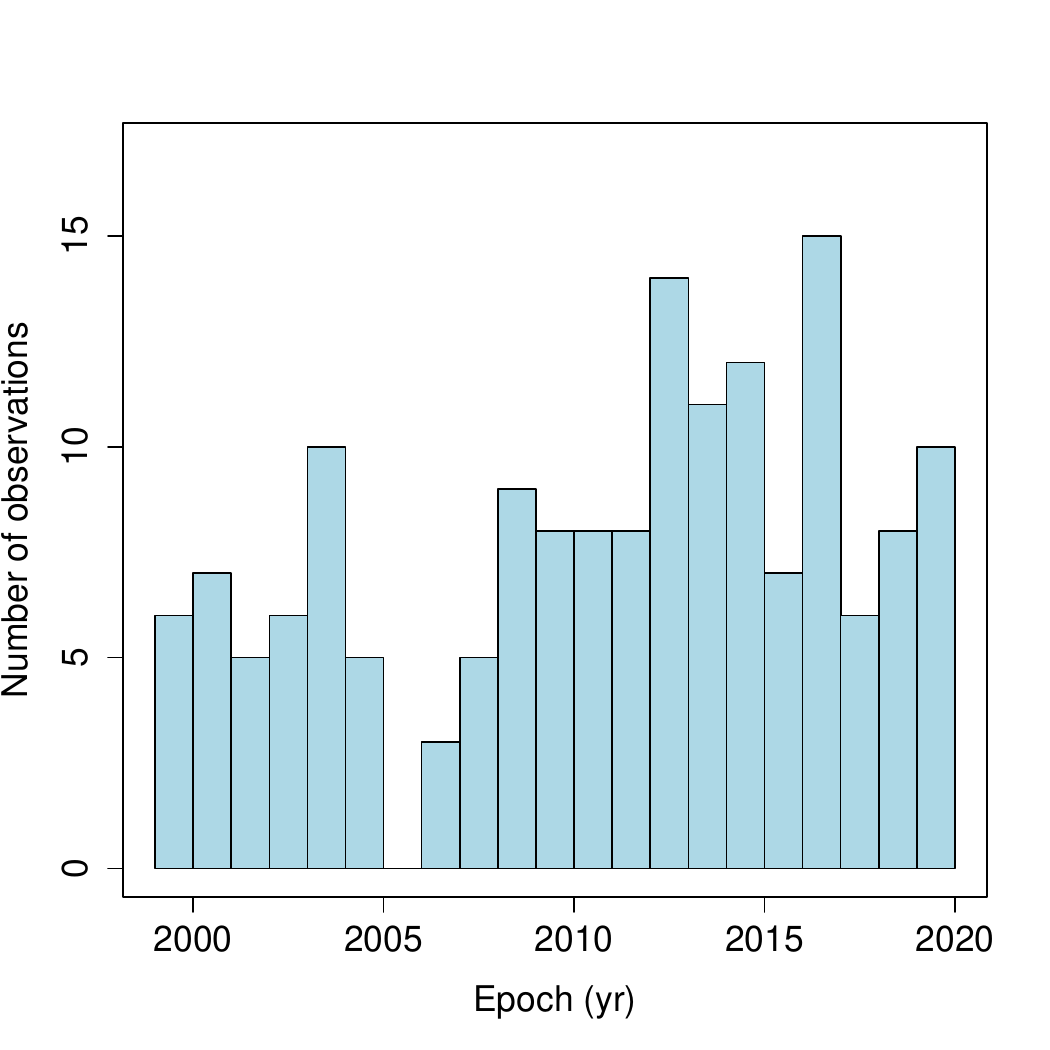}
\caption{Distribution of 164 observational epochs.} 
\label{fig:hist_obs_epochs}
\end{figure}

We use 172 epochs of VLBA observations of BL Lacertae between 1999.37 and 2019.97 made under the MOJAVE program and 2~cm VLBA survey \citep{lister09,kellermann98}. Data reduction, imaging and model fit of visibility data are carried out as in \cite{arshakian20}. A quasi-stationary component is present in 165 epochs. One epoch is excluded because the component stands well apart from the main cluster of positions. The remaining 164 quasi-stationary components cluster within 0.1~mas at the position angle PA$\approx -169\degr$. The epochs of observations have a gap between 2004.8 and 2006.4 (Fig.~\ref{fig:hist_obs_epochs}), and the frequency of observations is lower before the gap and higher after 2006.4 by a factor of about 1.5.
\FloatBarrier 

\begin{figure}[hpt]
\centering
\includegraphics[width=9cm,angle=0] {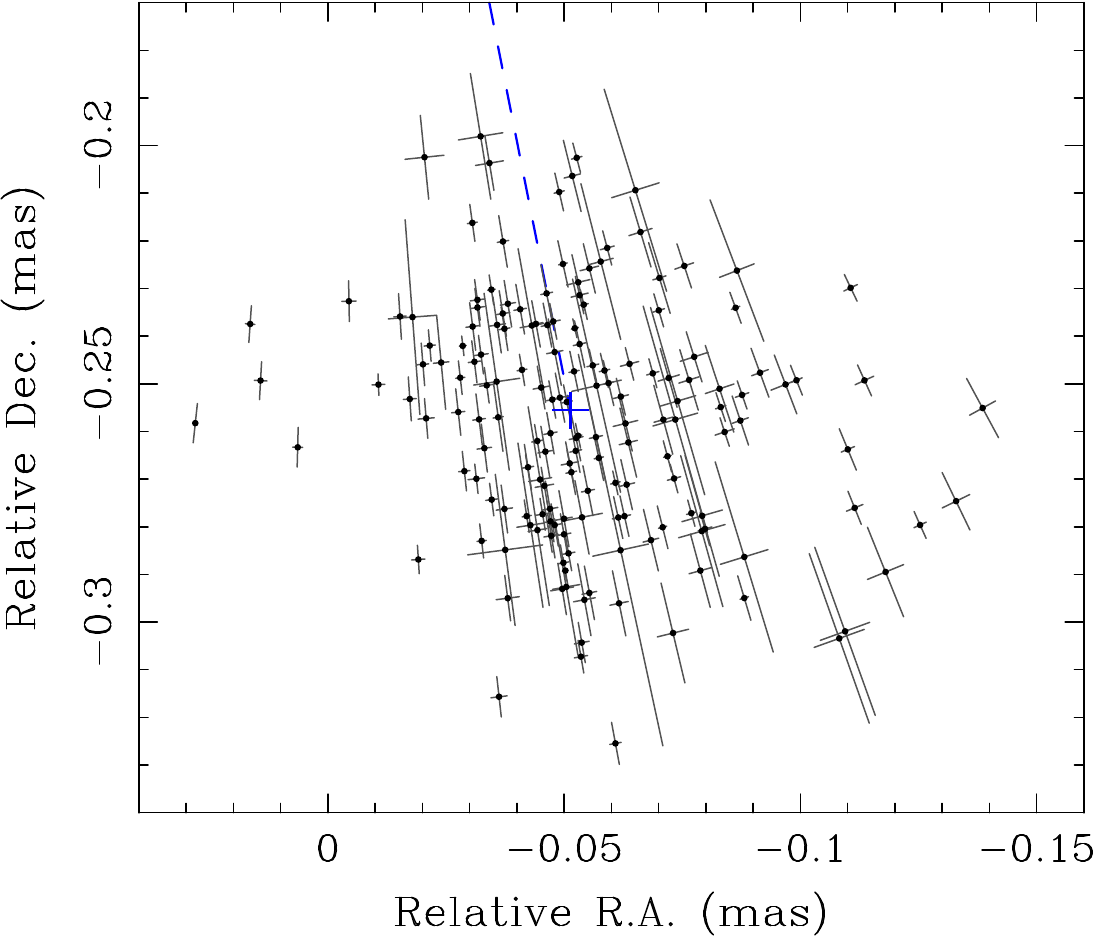}
\caption{Scatter of 164 positions of C7 on the sky plane. The sizes of the crosses correspond to C7 positional errors in the directions toward and across the core. The median position of the scatter is marked by a blue plus sign. The blue dashed line connects the median position of C7 and core.} 
\label{fig:c7_scatter_errors}
\end{figure}

The positions of C7 on the sky and their positional errors are shown in Fig.~\ref{fig:c7_scatter_errors}. Positional errors are estimated in directions along the positional angle relative to the core and transverse to it. The estimates of positional errors are obtained as in \cite{arshakian20} using the procedure suggested by \cite{lampton76} and based on minimising of the $\chi^2$ statistics. The median standard errors of C7 positions along and across the jet axis are $\delta_{\rm j} \approx 4.8~\mu$as and $\delta_{\rm n} \approx 1.5~\mu$as, respectively. The measured errors are treated as lower limits, while true errors can be higher by factor of few. They also estimated the flux leakage effect between the radio core and C7 and concluded that it is typically small, within 10\,\%. Throughout the paper, we assume that the measured positional uncertainties are close to proper errors.  

We verified this assumption by using the AIPS task UVMOD \citep{greisen03} to create twenty simulated VLBA epochs using homogeneous sphere models based on real BL Lacertae structure and noise levels, randomized for (u,v)-coverage and epoch.  The simulations were modelfit by another member of our team who did not know the location of C7 in each simulation. Comparison of the fitted location to known simulated C7 positions revealed standard deviations of $5.0~\mu$as and $2.7~\mu$as in the declination and right-ascension directions respectively, very similar to the estimated uncertainties given above.

\FloatBarrier 

\section{Trajectory analysis of quasi-stationary component.}
\label{sec:C7 trajectory analysis}
The compact radio core of the jet is usually the brightest feature when analysing images from VLBA observations and is used as a reference point for measuring the angular distances to the radio components of the jet. Measuring the distance between the core and the quasi-stationary component of C7 at different epochs allows us to study the dynamics of C7 and for this purpose \cite{arshakian20} introduced the apparent displacement vector $\vec{r}$, which defines the direction of motion of C7 between two consecutive epochs $t$ and $t+\Delta t$ and the length of the displacement, $r=\left|{\vec{r}}\right|$. If the radio core has proper motion, the measured apparent displacement C7 is a combination of the radio core displacement ($c$) and the C7 proper motion ($s$). The distributions of apparent displacements and their errors $\delta_r$, which are calculated by propagating the positional uncertainties of C7 at two consecutive epochs, are similar to those in \cite{arshakian20}. The median displacement and its mean error are 0.028~mas and 0.008~mas.

\cite{arshakian20} found that the larger displacement vectors have a tendency to be aligned along the jet axis. They interpreted it as evidence of wiggling of the radio core along the jet axis due to changes of physical conditions in the core region, which in turn scatters the intrinsic positions of C7 along the jet axis, and developed a method to estimate the intrinsic mean and standard deviation of the core and C7 displacements. The mean displacement of the core is $\overline{c}\approx 0$ and standard deviation $\sigma_{\rm c}=0.025$~mas, while the intrinsic motion of the C7 has a mean value $\overline{s} = 0.02$~mas and $\sigma_{\rm s} = 0.012$~mas. They concluded that both, anisotropic motion of the core and intrinsic motion of C7, have comparable contribution to the apparent motion of C7. Here we assume that the apparent motion of C7 is mainly due to the anisotropic motion of the core and the intrinsic motion of C7.

\subsection{Optimal time interval of a sliding window for the moving average procedure}
\begin{figure}[hpt]
\centering
\includegraphics[width=9.cm,angle=0] {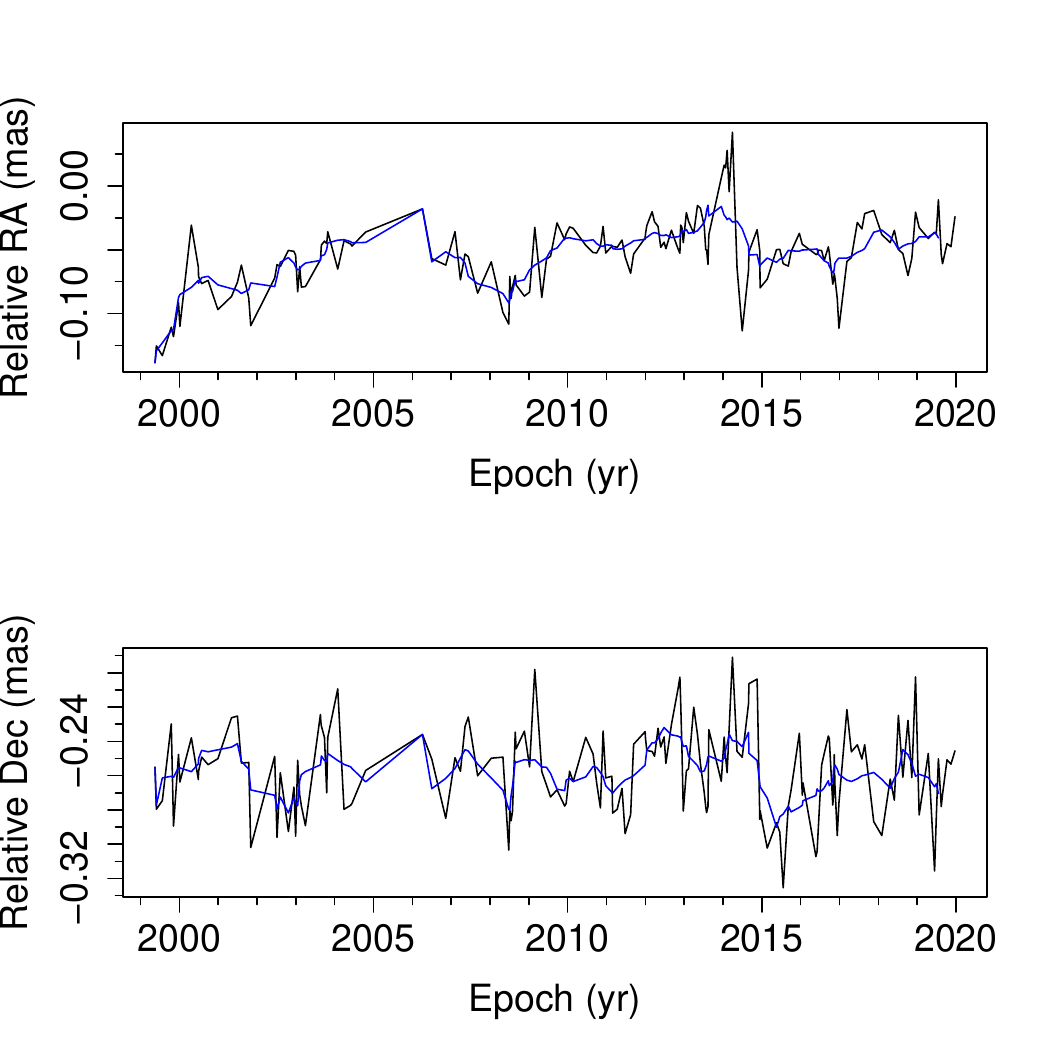}
\caption{Change of RA and Dec with epoch (black line). Smoothing with moving average procedure using the fixed time interval of $1$ yr (blue line). } 
\label{fig:RA-Dec_vs_epoch}
\end{figure}

To study the dynamics of the C7 intrinsic motion, we need to reduce the contribution of anisotropic core displacements, which have been shown to occur along the jet axis \citep{arshakian20}, and treat the asymmetric errors of the C7 positions. We use a moving average method that averages the anisotropic random motion of the radio core and any random C7 motions over a given time interval. For each epoch, the moving average procedure computes the average position of C7 and its error over a given time interval. The choice of the time interval of the sliding window should be optimal to average out the effects of the core displacement and at the same time preserve as far as possible the intrinsic trajectories of C7. For this purpose, we analyse the changes in the declination of C7 with time. The effect of the core displacement should be stronger along the Dec axis than along the RA axis (Fig.~\ref{fig:RA-Dec_vs_epoch}), because the direction of the central axis of the jet (PA$_{\rm jet} = -169\degr$), along which the core wiggles, almost coincides with the Dec axis. Indeed, change of position along Dec axis (bottom panel) shows stronger variation on scales of observing time intervals. The width of a sliding window should include several ups and downs for a reasonable smoothing of core displacements. Otherwise, the smoothing curve will follow the ups and downs which means that the core effect is not well smoothed. 
The quasi-oscillatory changes of the C7's position between 2015.5-2019 happen on time scales of about 1 yr (Fig.~\ref{fig:RA-Dec_vs_epoch}, bottom panel). It is unclear whether these changes are due to the effect of core displacement and/or intrinsic changes in the C7 position. To properly smooth these quasi-oscillatory changes, the averaging time interval should be on the order of about one year - the time scale of the quasi-oscillatory changes.

Another estimate of the time scales of cores displacements comes from consideration of ejection rate of moving components. Core displacements happen due to particle density and magnetic fields changes as well as brightening of the core due to passage of moving radio components through the core, as the region where opacity reaches $\tau=1$. For the BL Lac the latter happens about once per year \citep{lister21}. If the moving component is brighter than the core (e.g., as a result of the injection of denser plasma into the jet stream \citep{plavin19}), the inner region of the jet in the VLBI (Very Long Baseline Interferometry) image becomes elongated downstream with the brighter head displaced from the position of the core, resulting in an apparent displacement of the core. It suggests that the optimal smoothing time interval of $\Delta t_{\rm opt} = 1$ yr is a reasonable choice indeed.          

In the BL~Lac case, the propagating density perturbation of the jet plasma reaches the recollimation region (C7) from the core at 15~GHz in about four months \citep{arshakian20}, and its passage through the standing shock wave can affect its apparent motion in the direction of the jet axis both upstream and downstream. The time scale of these displacements should be comparable to the time scale of the core displacements, i.e., on the order of one year or less. Therefore, we believe that a smoothing time interval of about one year will smooth out the apparent motion of C7 along the jet axis.

Another potential bias in the fitted position of the components, which could lead to apparent regular changes in the position of C7, could be the result of model fitting, where we feed up the source model taken from the previous epoch for the current one and let it relax. However, this approach is justified for the model fitting adopted in the MOJAVE program data analysis, as (i) every source in the MOJAVE sample has its individual observing cadence according to the rate of its morphological changes derived from initial observing epochs, (ii) BL Lacertae is a record holder in terms of a number of observing epochs over all sources of the entire MOJAVE sample, with the highest cadence of one month. For a more detailed discussion of data reduction and model fitting, we refer the reader to \cite{lister09,lister18}.

 \FloatBarrier 
 \begin{figure*}[hpt]
 \centering
 \includegraphics[width=18cm]
 {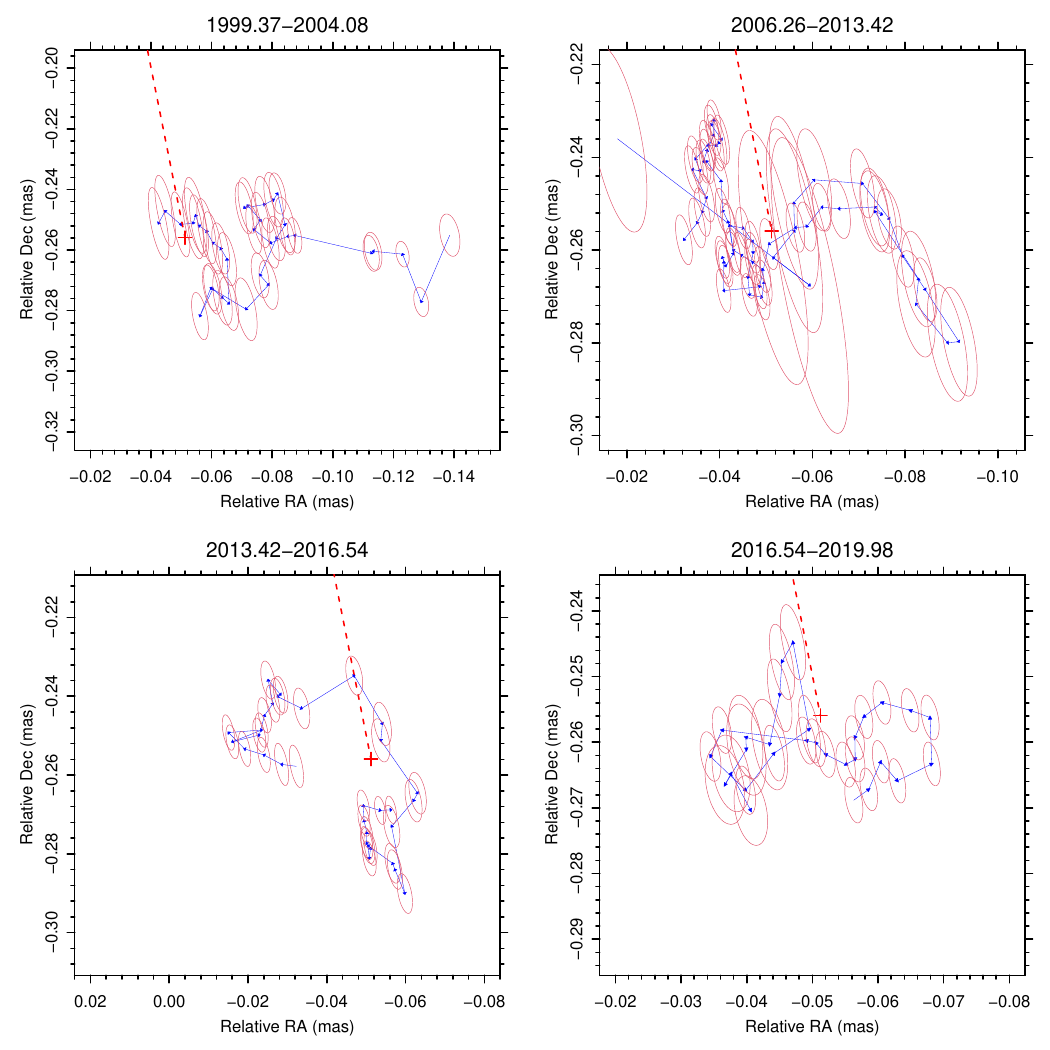}
 \caption{The trajectory of C7 smoothed by the moving average procedure on time scale of 1~yr (blue line). For clarity, the smoothed trajectory is represented for four time periods: 1999.37-2004.08, 2006.26-2013.42, 2013.42-2016.61 and 2016.61-2019.98. Average asymmetric positioning errors are shown by ellipses. The median scatter position over the whole time range from 1999.37 to 2019.98 is marked with a red plus sign. The red dashed line is the central axis of the jet connecting the median position of C7 and the radio core.}
 \label{fig:MA_dtau=1}
 \end{figure*}
%
%
%

\subsection{Smoothing the trajectory on time scale of one year}
The trajectory of C7 smoothed by moving average on time scales of about a year is shown in Fig. \ref{fig:MA_dtau=1} for four time periods. The average number of epochs in a rolling window of one year is about nine. The moving average procedure reduces the number of epochs to 155. The mean position error is asymmetric and it is represented as an ellipse (Fig.~\ref{fig:MA_dtau=1}) with the major axis directed to the core. The asymmetric $1\sigma$ position errors are represented by half of the major and minor axes.

\begin{figure*}[hpt]
\centering
\includegraphics[width=18cm] 
{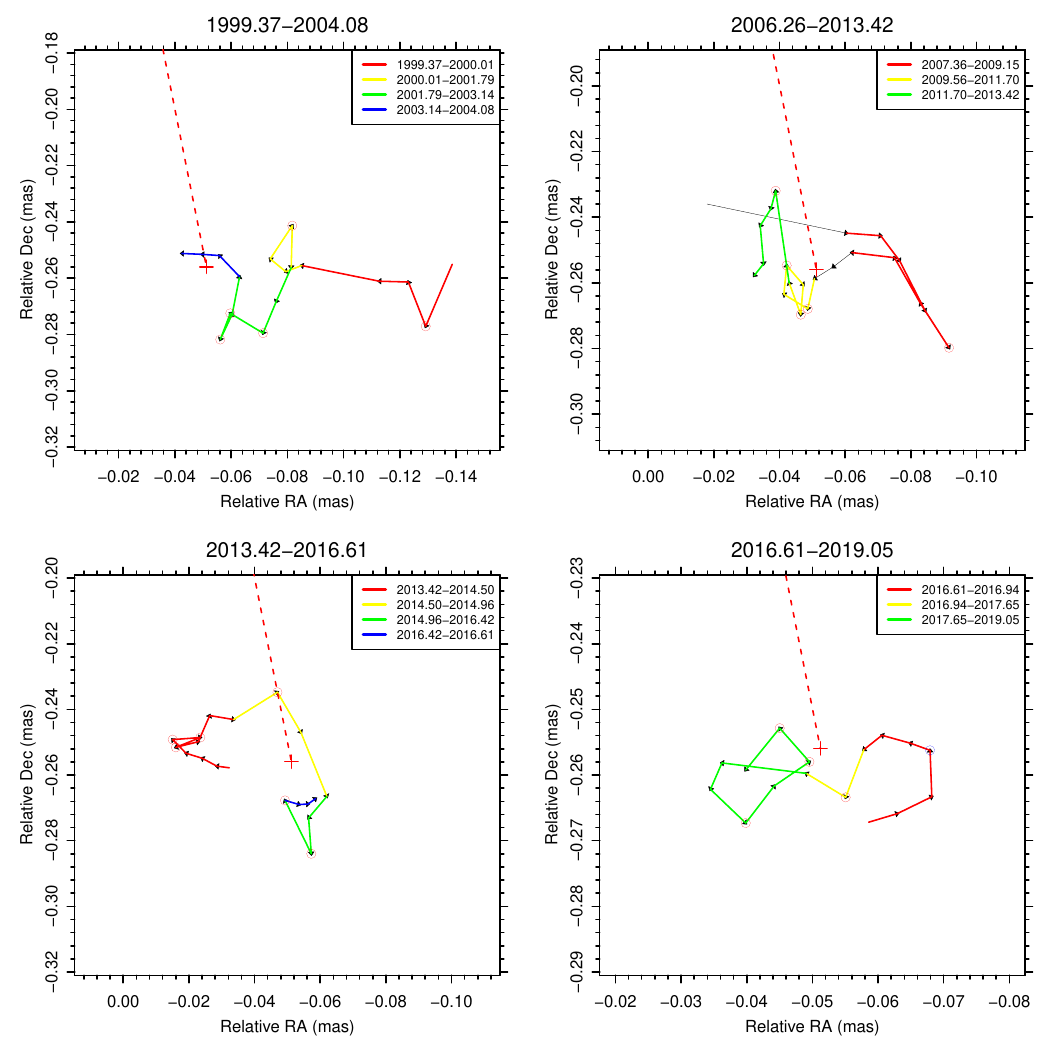} 
\caption{The refined trajectory of C7 is shown for four time intervals: 1999.37-2004.08, 2006.26-2013.42, 2013.42-2016.61 and 2016.61-2019.05. The arrows indicate the direction of movement. Red circles indicate reversal points for R- and RL-type, blue circles for A-type. Colored sections of the trajectory indicate the path lengths of identified reversals and their combinations. The median scatter position over the whole time range from 1999.37 to 2019.98 is marked with a red plus sign. The red dashed line is the jet central axis, which connects the median position of C7 and the radio core.} 
\label{fig:MA_dtau=1_clean}
\end{figure*}

It is evident that ellipse errors are intersecting along the displacement when it is small, which means that estimates of small displacements can be unreliable. For this we assume that the measurement of displacement is unreliable if the error ellipses of two consecutive positions are intersecting along the displacement vector or $1\sigma_i + 1\sigma_{i+1} > r_i$. The algorithm based on this criterion compares the error ellipses between two consecutive positions. If the displacement between two consecutive positions is less than the sum of their $1\sigma_i$ and $1\sigma_{i+1}$ positioning errors along the displacement (when the error ellipses intersect), the position with the larger error is discarded, and the smaller position error is compared to the ellipse error of the next position, otherwise, if $1\sigma_i + 1\sigma_{i+1} \le r_i$ then the ellipse errors $\sigma_{i+1}$ and $\sigma_{i+2}$ of the next two consecutive positions are compared. Application of this algorithm to the smoothed trajectory filters out about 60\,\% of positions. The resulting refined trajectory of C7 is shown in Fig.~\ref{fig:MA_dtau=1_clean}. 

\FloatBarrier 

We now carry out a visual inspection of the refined trajectory. The component C7 starts in 1999.37 far from the median center (plus sign) on the east side towards the median centre (red line), then makes a loop-like reversal at 2001.50 (yellow line) and moves to the south and sharply reverses at 2002.81 (green line). Notable is the long arcing movement to the northwest between 2002.81 and 2004.08 (yellow and green lines). There is a large gap of observations between 2004.80 and 2006.26. After 2007.36 the C7 moves in a long bending trajectory to the south-east and makes another sharp reversal at 2008.48 (red line). It shows a quasi-oscillatory clockwise motion between 2009.56-2011.70 (yellow line) and sharp backward motion in 2012.48 (green line).
Remarkably, starting around 2013.42 the C7 begins its distinctive clockwise bending away from the centre in a westerly direction and then back towards the centre (red line) and reverses at 2014.66 to the south along the jet axis (yellow line). Also noteworthy is the large arc-like trajectory from 2016.61 to 2016.94 (red line), which reverses shortly before and makes counterclockwise near oscillatory motion with an elongated orbit until late 2019 (green line). 

The whole trajectory of C7 motion can be represented as a combination of reversible and arc-like motions on various time (or spatial) scales.

\subsection{Reverse motion}
The first step in analysing of C7 motion is to check its randomness. If the motion of C7 is random, we should expect that the number of angles between successive displacements with $>90\degr$ should be approximately equal to the number of angles with $<90\degr$, i.e. their ratio $n(>90\degr)/n(<90\degr) \approx 1$ for a randomly directed displacements. This ratio is actually 2.5 times larger for angles estimated from the refined trajectory, $48/19 = 2.5$. This suggests that the movement of C7 is not random, it is rather directional, with alternating turns or reverse motion.

Another approach is to estimate the random probability of directional motion of C7: the probability that a sequence of $k$ random displacement vectors lies within a given angle $\phi$. We simulate 160 displacements with amplitude equal to one and random orientation from 0 to $2\pi$, and estimate the number of sequences of length $k$. We repeat this simulation 1000 times and estimate the average number of sequences of displacements $\overline{n_k}$. For example, we find that there are $\overline{n_3} = 3.46\pm0.06 \sim 4$ displacement sequences with three consecutive random displacements ($k=3$) that are aligned within $\phi = 60\degr$. The number of displacement sequences in the smoothed trajectory is $n_3 = 13$, which is about three times larger and indicates that parts of the C7 trajectory are not random but rather directional in nature.

The general reverse motion is the key to understanding the nature of C7 movement. 
Persistent reverse patterns are present on various spatial scales. To characterize it, we distinguish between a reversible trajectory (R) with an angle $\varepsilon \leq 90\degr$ at the turning point and an arc-shaped reversal (RA), which has no inner turning point ($\varepsilon > 90\degr$). If an R-type motion forms a loop, we call it RL-type. 

We define the time at which the reversal occurs $\Delta t_{\rm R} = t_{\rm e} - t_{\rm s}$, where $t_s$ and $t_e$ are the start and end times of the reversal and the length of the reversal $\Delta l_{\rm R}$ between $t_s$ and $t_e$. The epoch at the pivot point of the reversal is denoted by $t_{\rm R}$.

\begin{table}[htbp]
\caption{Characteristics of reverse motion.}
\label{table:1}
\centering
\begin{tabular}{cccc|cccc}
\hline
\multicolumn{4}{c|}{Reversals} & \multicolumn{4}{c}{Combo reversals} \\
\cline{1-4}\cline{5-8}
\multicolumn{1}{c}{} & \multicolumn{1}{c}{Type} & \multicolumn{1}{c}{Epoch} & \multicolumn{1}{c|}{cw/} & \multicolumn{1}{c}{} & \multicolumn{1}{c}{Type} & \multicolumn{1}{c}{Epoch} & \multicolumn{1}{c}{cw/}  \\
\multicolumn{1}{c}{} & \multicolumn{1}{c}{} & \multicolumn{1}{c}{(yr)} & \multicolumn{1}{c|}{ccw} & \multicolumn{1}{c}{} & \multicolumn{1}{c}{} & \multicolumn{1}{c}{(yr)} & \multicolumn{1}{c}{ccw}  \\
\hline
1 & R & 1999.41 & cw & & & & \\
2 & RL & 2001.50 & cw & & & & \\
\rowcolor[gray]{0.95}
3 & R & 2002.52 & cw & & & & \\
\rowcolor[gray]{0.95}
4 & R & 2002.60 & ccw & 1 & RC & 2002.81$^{c}$ & ccw \\
\rowcolor[gray]{0.95}
5 & R & 2002.81 & ccw & & & & \\
6 & RA$^{a}$ & - & ccw & & & & \\
7 & R & 2008.48 & cw & & & & \\
\rowcolor[gray]{0.95}
8 & R & 2009.73 & cw & & & &  \\
\rowcolor[gray]{0.95}
9 & RL & 2010.65 & cw & 2 & RO & 2011.14$^{d}$ & cw \\
\rowcolor[gray]{0.95}
10 & R & 2011.14 & cw & & & & \\
11 & R & 2012.48 & ccw & & & & \\
\rowcolor[gray]{0.95}
12 & R & 2013.61 & cw & & & & \\
\rowcolor[gray]{0.95}
13 & R & 2013.64 & cw & 3 & RC & 2013.61$^{c}$ & cw \\
\rowcolor[gray]{0.95}
14 & R & 2013.96 & ccw & & & & \\
15 & R & 2014.66 & cw & & & & \\
16 & R & 2015.46 & cw & & & & \\
17 & R & 2016.42 & cw & & & & \\
18 & RA & 2016.85$^{b}$ & ccw & & & & \\
19 & R & 2017.31 & cw & & & & \\
\rowcolor[gray]{0.95}
20 & RL & 2018.31 & ccw & & & & \\
\rowcolor[gray]{0.95}
21 & R & 2018.52 & ccw & 4 & RO & 2018.31$^{d}$ & ccw \\
\rowcolor[gray]{0.95}
22 & R & 2018.86 & ccw & & & &  \\
\hline
\end{tabular}
\tablefoot{Two main columns separate the characteristics of reversals and combination of reversals. The columns for reversals are as follows: Reversal type 
and direction of reverse motion. The columns for combination of reversals are the type, epoch and direction of reverse motion.  \\
\tablefoottext{a}{The type of reversal is questionable because of the proximity of the smoothed trajectory to the gap in the observational data.}  \\
\tablefoottext{b}{The epoch 2016.85 (blue circle in Fig.~\ref{fig:MA_dtau=1_clean}) of the C7 position at the maximum distance from the starting position of the reversal at epoch 2016.61 (see Fig.~\ref{fig:MA_dtau=1_clean} and Table~\ref{table:2}) is taken as the epoch of the turning point for the RA-type reversal.} \\
\tablefoottext{c}{The epoch of the C7 position at the maximum distance from the starting position of the reversal (see Table~\ref{table:2}) is taken as the epoch of the turning point for the RC-type reversal.} \\
\tablefoottext{d}{The epoch of the turning point of the quasi-oscillatory reversal is taken as the epoch of the top of the reversal, which is closest to the epoch corresponding to $t_{\rm s} + (t_{\rm e}-t_{\rm s})/2$.} 
}
\end{table}


In Table~\ref{table:1}, we give 22 identifications of C7 trajectories. There are 21 reversing patterns with pivot points marked by circles in Fig.~\ref{fig:MA_dtau=1_clean}. Among them, 17 are R-type, 3 RL- and 1 RA-types (see Table~\ref{table:1}). One trajectory between 2003.14-2004.08 (blue line) is identified as probable RA type reversal with unknown epoch of a turning point. Most of reversals is of R-type. For clarity, we choose the likely epochs for the beginning and end of each reversal, $t_{\rm s}$ and $t_{\rm e}$, and present the trajectory of the reversals in different colours (Fig.~\ref{fig:MA_dtau=1_clean}).
A clear R-shaped trajectory reversal at spatial scales of order 0.05-0.07~mas is observed between $2007.36-2009.15$ (red line) and $2014.50-2014.96$ (yellow line). We believe that a reversal of the same order occurs in an easterly direction with a pivot point at $1999.41$ (red line), but we only observe its backward motion towards the median centre of C7 positions. There are 14 more R-type reversals on scales from about 0.01 to 0.05~mas (Table~\ref{table:2}). We suppose that reversals on smaller scales are there but the scale of the smoothing interval (1 yr) and cleaning procedure puts a lower limit on visibility of small reversals. A remarkable example of an arc-shaped reversal (RA-type) is observed between 2016.54 and 2016.94 (red line). Here we assume that the reversal pivot point is at 2016.85, whose position is at the maximum distance from the starting position of the reversal. It is possible that a reversal of the same type occurs after 2003.14 (blue line), but the paucity of data at the threshold of the observational gap does not allow us to draw a decisive conclusion.   

Among the 22 reversals (Table~\ref{table:1}), 13 are clockwise (CW) and 9 are counterclockwise (CCW). This difference is statistically insignificant, and we assume that the direction of rotation of reversals is random. 

\begin{figure}[ht]
\centering
 \includegraphics [width=7cm]{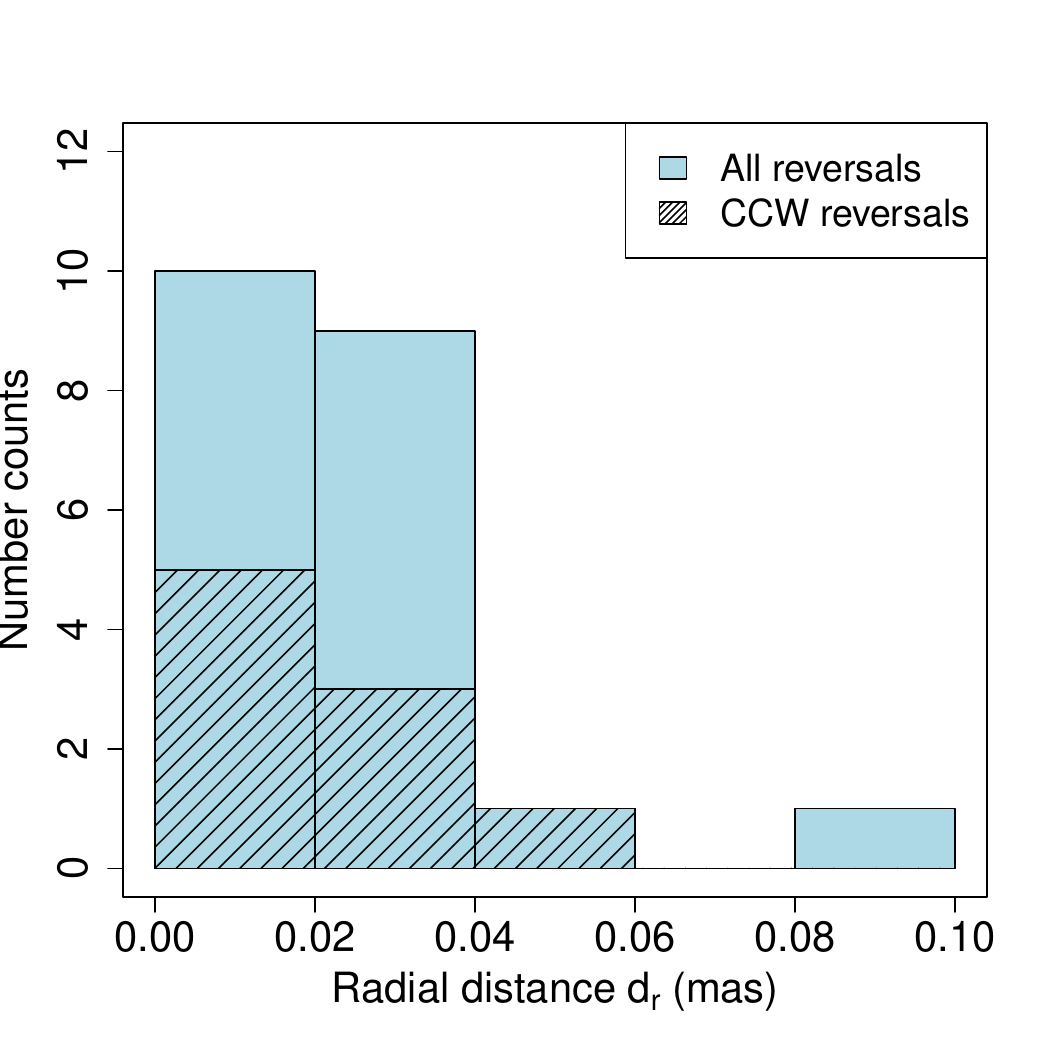}
  \caption{Distribution of radial distances of 21 turning points from the median position of the C7 scatter. The shaded area shows the distribution of radial distances of the turning points of reversals with CCW motion.}
\label{fig:hist_rdist}
\end{figure}

We measure the radial distances $d_{\rm r}$ between the pivot points of reversals and the median center of the scatter of C7 positions. The radial distances of the reversal points are mainly in the range of 0.003~mas to 0.037~mas with an average radial distance of about $0.025 \pm 0.004$~mas (see Fig.~\ref{fig:hist_rdist}). The distribution of radial distances to the reversal points with CCW motion (shaded area) is not statistically different as compared from those with CW motion. The azimuthal angles of the turning points are measured with respect to the jet axis (counterclockwise). There is a lack of turning points between $90\degr$ and $180\degr$, which may be due to the small number of angles statistics. The CW and CCW reversals have similar distributions, that is, their pivot points occur in random directions.
\begin{figure}[ht]
\centering
 \includegraphics [width=7cm]{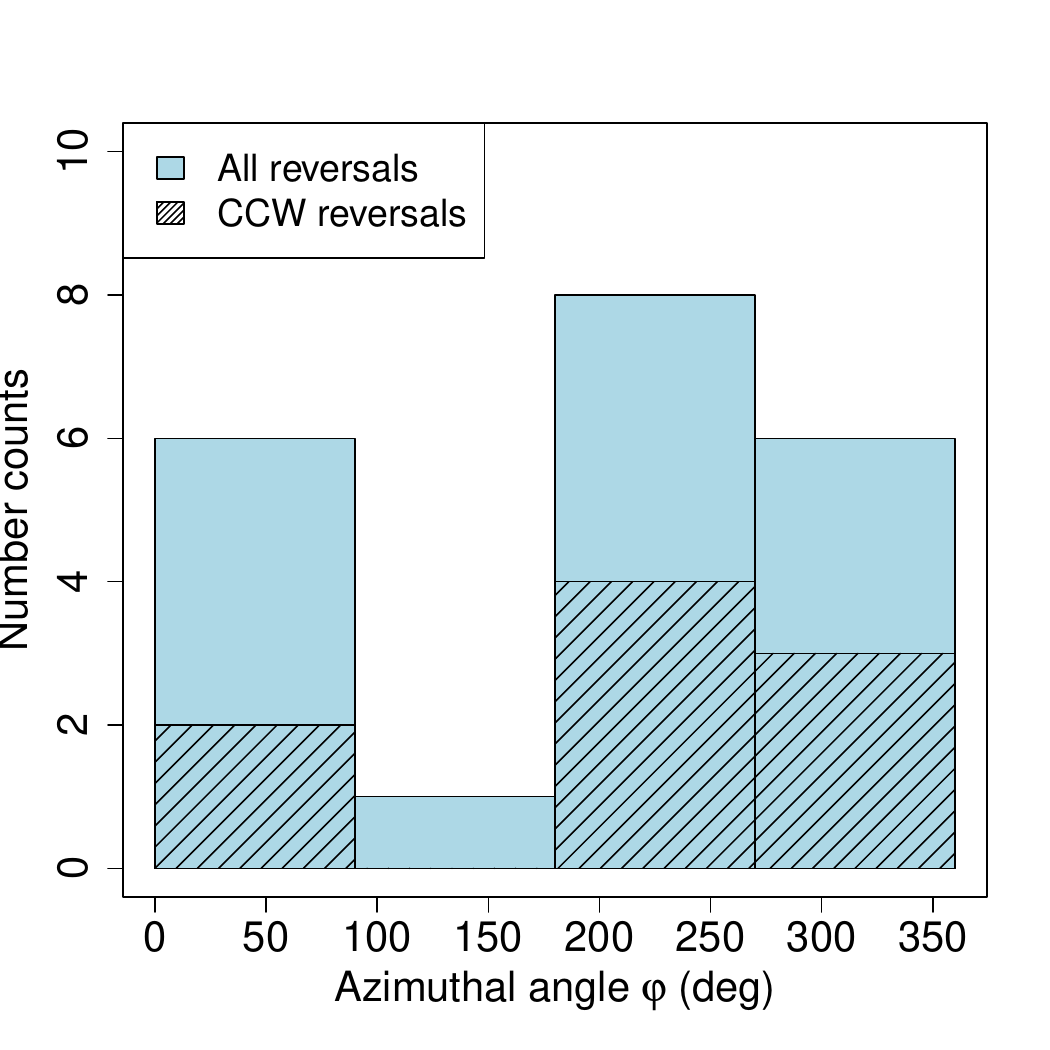}
  \caption{Distribution of azimuthal angles for 21 turning points. The shaded area shows the distribution of azimuthal angles of the turning points of reversals with CCW motion.}
\label{fig:hist_azimuth}
\end{figure}

We measure 18 time intervals between pivots of successive reversals $\Delta \tau$ (three time intervals are excluded from the statistics: one time interval due to uncertain epoch of the turning point of the RA-type reversal and two time intervals between 2002.81 and 2008.48 due to the  observation gap). They have a fairly flat distribution within the range of 0.03-1.34 yr and have a mean value of $\overline{\Delta\tau} = 0.65\pm0.10$ yr, which corresponds to a reversal frequency of $f_{\rm R}\approx 1.5$ yr$^{-1}$.


\paragraph{Combination of reversals.} We observe reversible trajectories consisting of a combination of smaller reversals (or sub-reversals). For example, three consecutive reversals, two R-type with vertices at 2002.52 and 2002.60 and one RL 
reversal at 2002.81 (three red circles in Fig.~\ref{fig:MA_dtau=1_clean}; Table~\ref{table:1}) have two common displacements, spatial scales of about 0.012, 0.011, and 0.008~mas, respectively, and two of them have reversal angles close to 
$90\degr$. We assume that all three reversals are part of a large reversal occurring between 2001.79 and 2003.14 (see the green line in 
Fig.~\ref{fig:MA_dtau=1_clean} and Table~\ref{table:2}) with a turning point at 
2002.81 which is the furthest from the initial onset of the reversal. Three 
analogous small-scale reversals, manifesting within spatial scales below 0.01~mas, 
contribute to a large R-type reversal observed between 2013.42 and 2014.50. The latter comprises two R-type and one RL-type reversals, occurring respectively at 2013.61, 2013.64, and 2013.96 (red circles). Notably, 2013.61 is the furthest reversal point from the initial onset, thus being designated as the primary pivot point (Fig.~\ref{fig:MA_dtau=1_clean}; Table~\ref{table:1}). We call a reversal consisting of a combination of sub-reversals an RC-type.
\begin{table*}[htbp]
\caption{Characteristics of reversals and reversal combinations. The latter are marked with a gray bar.}
\label{table:2}
\centering
\begin{tabular}{cccccccc}
\hline
\multicolumn{1}{c}{} & \multicolumn{1}{c}{Type} & \multicolumn{1}{c}{Start} & \multicolumn{1}{c}{End} & \multicolumn{1}{c}{Time interval} & \multicolumn{1}{c}{Length} & \multicolumn{1}{c}{Speed} & \multicolumn{1}{c}{Spatial scale}\\
\multicolumn{1}{c}{} & \multicolumn{1}{c}{} & \multicolumn{1}{c}{(yr)} & \multicolumn{1}{c}{(yr)} & \multicolumn{1}{c}{(yr)} & \multicolumn{1}{c}{(mas)} & \multicolumn{1}{c}{(c)} & \multicolumn{1}{c}{(mas)}\\
\hline
1 & R$^{a}$ & 1999.37 & 2000.01 & 0.64 $(\times2)$ & 0.08 $(\times2)$ & 0.13 & 0.05 \\
2 & RL & 2000.01 & 2001.79 & 1.78 & 0.04 & 0.10 & 0.02 \\
\rowcolor[gray]{0.95}
3 & RC & 2001.79 & 2003.14 & 1.35 & 0.08 & 0.24 & 0.04 \\
4 & RA & 2003.14 & 2004.08 & 0.94 & 0.06 & 0.29 & 0.04 \\
5 & R & 2007.36 & 2009.15 & 1.79 & 0.08 & 0.18 & 0.05 \\
\rowcolor[gray]{0.95}
6 & RO & 2009.56 & 2011.70 & 2.14 & 0.05 & 0.10 & 0.01 \\
7 & R & 2011.70 & 2013.42 & 1.72 & 0.06 & 0.14 & 0.03 \\
\rowcolor[gray]{0.95}
8 & RC & 2013.42 & 2014.50 & 1.08 & 0.06 & 0.24 & 0.02 \\
9 & R & 2014.50 & 2014.96 & 0.46 & 0.05 & 0.46 & 0.04 \\
10 & R & 2014.96 & 2015.96 & 1 & 0.03 & 0.12 & 0.02 \\
11 & R & 2015.96 & 2016.61 & 0.65 & 0.02 & 0.13 & 0.01 \\
12 & RA & 2016.61 & 2016.94 & 0.33 & 0.06 & 0.77 & 0.03 \\
13 & R & 2016.94 & 2017.65 & 0.71 & 0.03 & 0.17 & 0.02 \\
\rowcolor[gray]{0.95}
14 & RO & 2017.65 & 2019.05 & 1.4 & 0.11 & 0.32 & 0.03 \\
\hline
\end{tabular}
\tablefoot{The columns for reversals are as follows: Reversal type, the epochs of the beginning and end of the reversal, $t_{\rm s}$ and $t_{\rm e}$, reversal time interval $\Delta t_{\rm R} = t_{\rm e} - t_{\rm s}$, length of the reversal $\Delta l_{\rm R}$, speed of the C7 component $\beta^*_{\rm s}$ and spatial scale which represents the largest size of the reversal. \\
\tablefoottext{a}{Only part of the reversal is observed. The length and time interval of the reversal are taken to be equal to twice the measured length and time interval.}
}
\end{table*}
We identify another type of reversal combination, such as the trajectory between 2009.56 and 2011.70 (yellow line), which exhibits a quasi-oscillatory motion consisting of two R-type reversals and one RL-type reversal on a small spatial scale of 0.012~mas (Table~\ref{table:2}). This is consistent with the small amplitude of the transverse jet wave ($\approx 0.02$~mas) during the period of low jet activity between 2010 and 2012 \citep{cohen15,arshakian20}. A similar quasi-oscillatory trajectory is observed between 2017.65 and 2019.05 (green line) on time scales of about 0.03~mas, which consists of one RL- and two R-type reversals at turning points 2018.52 and 2018.86. The combination of reversals exhibiting quasi-oscillatory motion is labelled as RO-type. Notably, RO-reversals consist of sub-reversals with the same direction of motion, CW or CCW (see Table~\ref{table:1}).

We estimate the apparent speeds of C7 during reverse motion as $\beta_{\rm s}^* = \Delta l_{\rm R} / \Delta t_{\rm R}$, which range from $0.03\,c$ to $0.77\,c$ (Table~\ref{table:2}). These subluminal speeds are actually lower limits as the $\Delta l_{\rm R}$ distances decrease due to smoothing of the C7 trajectory on time scales of one year. For the same reason, the spatial scales of reversals, which are estimated as the maximum reversal size, represent lower limits. The highest speed ($0.77\,c$) of C7 is observed for the RA-type reversal (red line between 2016.61-2016.94 and Table~\ref{table:2}), while the lowest speed of $0.03\,c$ is estimated for the R-type reversal. It is noteworthy that both reversals occur at a spatial scale of 0.03~mas. The average time interval of the R and RL reversals is about $1.1 \pm 0.2$~yr, which is about three times that of one RA type. As for the four combined reversals, the average time interval is about 1.2~yr and 1.8~yr for the two RC and two RO types, respectively. Additional statistics for RA, RC, and RO types are needed to confirm these data. We also compared the distributions of time intervals $\Delta t_{\rm R}$ for CW and CCW reversals and found no significant difference.  \\

To test whether the results of section 3 hold for a smoothing time interval longer than the optimal interval of one year, we repeated the analysis of this section for a smoothing time interval of 1.5 yr. The number of reversals decreased from 22 to 12 with an equal number of clockwise and counterclockwise movements (Table~1). The ratio characterising the directional motion of C7 becomes larger, $n(>90\degr)/n(<90\degr) = 38/12 \approx 3$, indicating that the movement of C7 is not random. All seven reversals on time scales of $\gtrsim 1.5$ years were preserved (Table~2). Among them there are three R-type, two RO-type, one RC and one RL-type in 2001.5, which changed to R-type after smoothing. The remaining reversals are R-type (or change to it) on time scales of $>0.3$ year. Thus, the C7 movement is directional, the reversals persist mainly on longer time scales, and the directions of the reversals are random. We conclude that the main results of this section remain unchanged.

\section{Model for reversible motion}
\label{sec:C7 model}
\cite{cohen15} reported a close correlation between C7 position angles and the jet ridge line. Based on this relationship, they suggested that C7 excites transverse waves that propagate downstream the jet with relativistic speeds. \cite{arshakian20} showed a correspondence between the spatial scales of C7 motion and the amplitudes of relativistic transverse waves. In this study, we have shown that C7 performs reversible motion on different spatial scales. This suggests that the reversible motion of the stationary component of C7 is a projection of its motion along the transverse wave of the jet. Here we assume that transverse waves are generated up to C7 upstream of the jet. In this case, the passage of the wave through the location of C7 causes it to make oscillatory motions across the jet, similar to the motion of a boat at anchor on a wave. Depending on the geometry of the wave, reversible motions of different types can be observed. Let us consider a simple case when the line of sight coincides with the central axis of the jet and the wave propagates in the plane passing through the line of sight. Then the motion of the wave along the axis of the jet will shift the stationary component C7 across the jet axis. If the wave is periodic then C7 will oscillate transverse to the axis and the observer will register a linear reversal motion of C7. Such a linear reversal motion can also be observed if the jet axis makes some angle with respect to the line of sight and the plane of the wave passes through the jet axis and the line of sight (or close to them). In this case, the direction of the linear reversal should be (or nearly) parallel to the jet axis, as for example in the case of the reversal occurring between 2007.36-2009.15 (Fig.~\ref{fig:MA_dtau=1_clean}, red line). In most cases, the reversal motions of C7 are not linear, indicating that the waves have a twist in 3-dimensional space, i.e., a swirled structure. If the wave is periodic and the twist angle is constant, C7 exhibits quasi-circular motion (as the yellow line between 2009.56-2011.70 and green line between 2017.65-2019.05) if the jet is viewed within the jet opening angle, and ellipse-like motion (e.g., green line in 2011.70-2013.42) when the jet viewing angle is larger than the jet opening angle. Certain combinations of wave twist angle, wave twist direction, and jet viewing angle can lead to looping reversals, such as the RL-type reversal observed during 2000.01-2001.79.

In this scenario, the passage of the wave through the C7 position moves the latter from the wave trough down the slope towards the crest, rounds the crest and then heads towards the wave trough. This movement of the C7 is observed as a reversible movement with a pivot point at the wave ridge position. Each observed reversal corresponds to a transverse wave passing through C7, hence, the mean transverse wave frequency $f_{\rm w} = f_{\rm R} = 1.5$~yr$^{-1}$ and mean wave period $T_{\rm w} = 1/f_{\rm R} = \overline{\Delta \tau} \approx 0.65 \pm 0.1$~yr. We define the wave amplitude $A_{\rm w}$ as the distances from the pivot point to the midpoint of the distance between the initial and final positions of the reversal. We use the measured spatial scales of the reversals (Table~\ref{table:2}) as a proxy for their amplitudes. There is a small deviation between the two for reversals with large angles at the turning points (see, e.g., Fig.~\ref{fig:MA_dtau=1_clean}, yellow line between 2014.50-2014.96), but for the other reversals the deviation is negligible. The most energetic waves with large amplitudes  $A_{\rm w} \sim 0.05$~mas belong to the R-type, and conversely, the weakest waves have small amplitudes on the order of 0.01~mas and appear as different types, R and RO. The speed of C7 in the reference frame of the host galaxy depends on the velocity of the transverse wave and wave's twist at the position of C7. In the next section, we detail the apparent speeds of C7 measured by the observer in the context of the proposed C7 model.

\section{Superluminal speeds of quasi-stationary component}
\label{sec:C7 superluminal speeds}
\begin{figure}[hpt]
    \centering
    \includegraphics[angle=0,width=\linewidth]{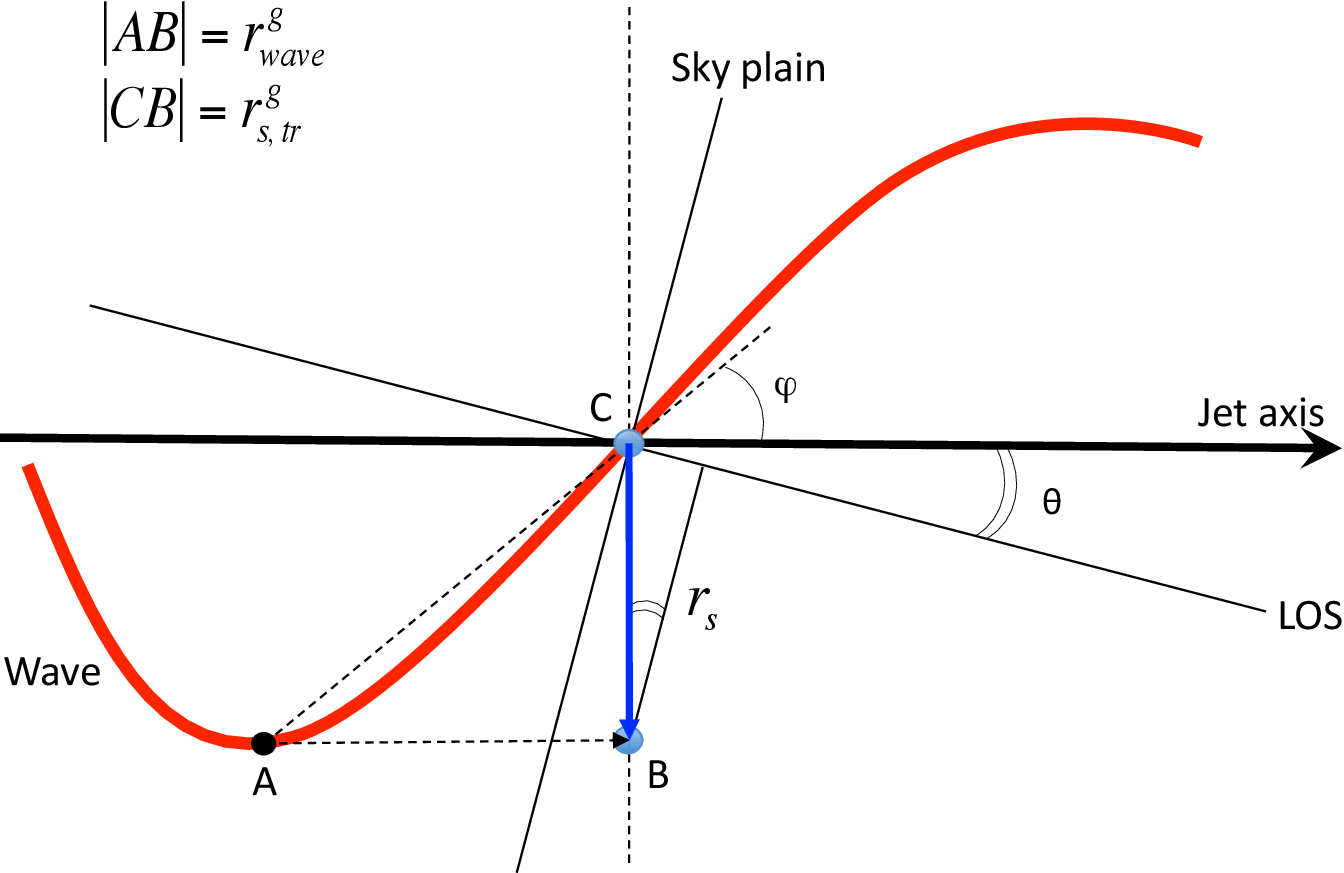}    
    \caption{Sketch of the apparent transverse motion of the quasi-stationary component C7 ($\overrightarrow{CB}$) as the relativistic transverse wave travels along the jet axis ($\overrightarrow{AB}$). }
    \label{fig:sketch}
\end{figure}

\cite{arshakian20} estimated 111 transverse speeds of C7. The mean apparent speed was found to be superluminal $\overline{\beta}_{\rm s} = 4.6\,c$, with $\approx 90\,\%$ of the speeds larger than the speed of the light (this fraction holds also for the 163 transverse speeds measured in this study). They attributed this to significant core displacements and used projections of the C7 displacements onto the axis transverse to the jet, which is unbiased (i.e., free of core displacement effects), to estimate the C7 apparent speeds and found that the lower limits of most speeds $\beta_{\rm s} > 1.2\,c$ still remain superluminal. It remained unclear why the speeds of the quasi-stationary component, which moves in a plane almost normal to the jet axis, should be superluminal.
To register superluminal speeds, there must be a relativistic component that moves at a small angle to the line of sight. We assume that the transverse wave of the jet moving with relativistic speed at small viewing angle shifts the quasi-stationary component in the transverse direction from the jet with subluminal speed, which in the observer's rest frame is registered as superluminal. For simplicity, assume that the wave propagates in the sagittal plane (the plane containing the LOS and the central axis of the jet; see Fig. 6 in Cohen et al. 2015). The sagittal plane is shown in Fig. \ref{fig:sketch} when viewed from a direction normal to the sagittal plane. 

Let us assume that the quasi-stationary component is located at the intersection of the transverse wave and the jet axis at point C. While the wave propagates along the jet central axis with relativistic speed $\beta^g_{\rm wave}$ (in the frame of a host galaxy) and passes the distance $|AB|=r^{g}_{\rm wave}$ in the longitudinal direction in a time interval ${\delta t}^{g}$, the C7 travels a distance $|CB|=r^{g}_{\rm s, tr}$ in the direction transverse to the jet axis with a speed $\beta^{g}_{\rm s, tr} = r^{g}_{\rm s, tr} / {\delta t}^{g}$. Let $\theta$ be the viewing angle of the jet along the central axis. An observer measures the displacement of C7 ($r_{\rm s} = r_{\rm s,tr}\cos{\theta}$) over time $\delta t$ and hence the apparent transverse velocity of C7, $\beta_{\rm s} = r_{\rm s}/\delta t$. Given that $r^{g}_{\rm s, tr} = r_{\rm s}/\cos\theta$ and relativistic time transformation ${\delta t}^{g} = \delta t/(1-\beta^{g}_{\rm wave}\cos\theta)$ we obtain an expression for the transverse velocity of C7 in the frame of a host galaxy,
\begin{equation}
    \beta^{g}_{\rm s, tr} = \frac{\beta_{\rm s}(1-\beta^{g}_{\rm wave}\cos\theta)}{\cos\theta}.
    \label{eq:beta^g_s}
\end{equation}
Consider the ratio 
\begin{equation}
    \frac{\beta^g_{\rm s,tr}}{\beta^g_{\rm wave}} = \frac{r^g_{\rm s,tr}}{r^g_{\rm wave}} = \tan{\varphi},  
    \label{eq:tg_varphi}
\end{equation}
where $\varphi$ is the wave twist angle, that is the angle between the wave tangent and the direction of the jet central axis, which ranges from $0\degr$ to $\varphi_{max}<90\degr$. Using the Eq.~(\ref{eq:tg_varphi}), we invert Eq.~(\ref{eq:beta^g_s}) to express the apparent transverse speed of C7  measured by observer,
\begin{equation}
    \beta_{\rm s} = \frac{\beta^{g}_{\rm wave}\tan\varphi \cos\theta}{(1-\beta^{g}_{\rm wave}\cos\theta)}.
    \label{eq:beta_s_sagittal}
\end{equation}
If the wave propagates in a plane perpendicular to the sagittal plane, the transverse motion vector of the C7 makes a right angle to the line of sight and remains so irrespective of changing the viewing angle. In this case $r^{g}_{\rm s, tr} = r_{\rm s}$ and the Eq.~(\ref{eq:beta_s_sagittal}) transforms to,
\begin{equation}
    \beta_{\rm s} = \frac{\beta^{g}_{\rm wave}\tan\varphi}{(1-\beta^{g}_{\rm wave}\cos\theta)}.
    \label{eq:beta_s_perp}
\end{equation}
The apparent speeds of the same waves propagating in the sagittal plane and normal to the sagittal plane (Eqs.~ (\ref{eq:beta_s_sagittal}) and (\ref{eq:beta_s_perp})) may differ substantially if the viewing angle is large. For small $\theta<10\degr$ \citep{cohen15,pushkarev17,homan21} the speed difference is negligible.  

\begin{figure}[hpt]
    \centering
    \includegraphics[angle=90,width=\linewidth]{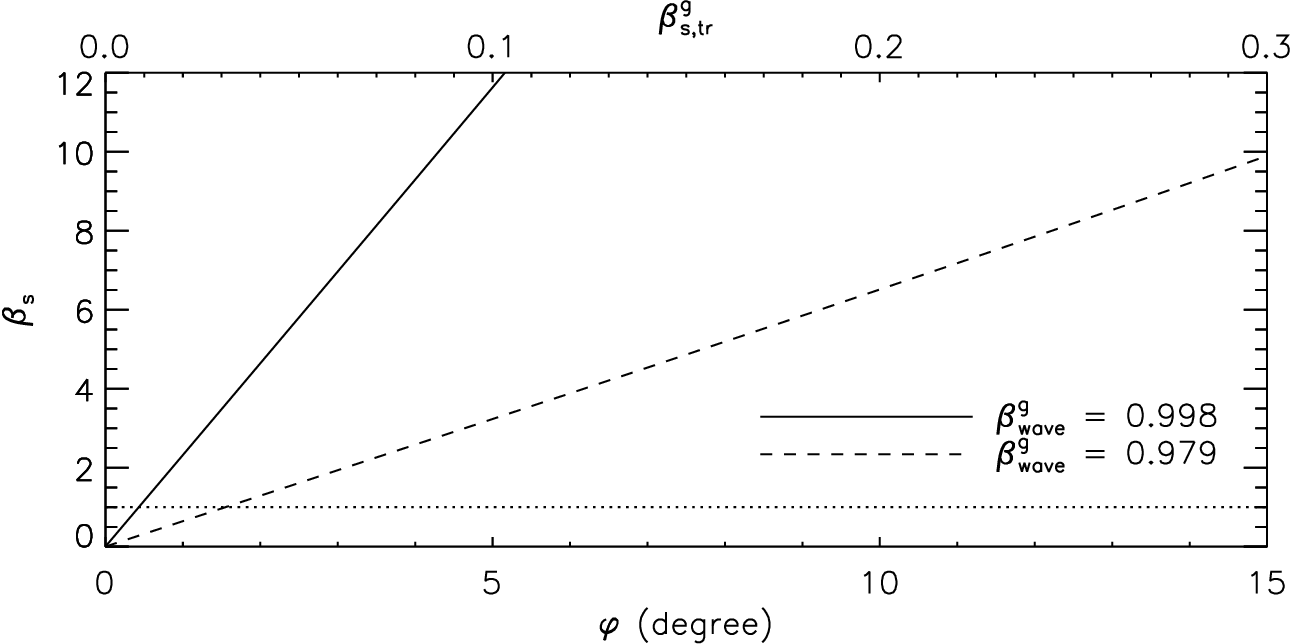}
    \caption{Apparent transverse velocity $\beta_{\rm s}$ of C7 as a function of wave twist angle $\varphi$ and transverse velocity $\beta^{g}_{\rm s}$ (in the frame of host galaxy) for a fixed viewing angle $\theta=6\degr$ and longitudinal speeds of the fastest and slowest transverse waves $\beta^{g}_{wave}=0.998$ and $\beta^{g}_{wave}=0.979$ (or the corresponding Lorentz factors, 15 and 3.5; full and dashed lines respectively). The dotted horizontal line separates subluminal and superluminal apparent speeds.}
    \label{fig:beta_s-phi}
\end{figure}

\paragraph{Transverse wave speeds.}
\cite{cohen15} estimated the longitudinal speeds of four transverse waves adopting $\theta = 6\degr$ (see their Table~1). They used the shift of ridge lines in time at a distance of about 2-3~mas from the radio core. All transverse waves have relativistic speeds and are in the range $\beta^{g}_{\rm wave} = 0.979-0.998$, which corresponds to the range of Lorentz factors $\Gamma_{\rm wave} = 3.5-15$. We assume that transverse waves have the same speeds at a time when passing C7. The change of $\beta_{\rm s}$ as a function of $\varphi$ and $\beta^{g}_{\rm s, tr}$ for the the fastest and slowest transverse waves, $\beta^{g}_{\rm wave} = 0.998$ and $\beta^{g}_{\rm wave} = 0.979$ (Fig. \ref{fig:beta_s-phi}, full and dashed lines).
The transverse speed $\beta^{g}_{\rm s}$ of C7 depends on the wave speed $\beta^{g}_{\rm wave}$ and its twist angle $\varphi$. The fastest wave (full line) with slopes $>0.5\degr$ moves C7 with subrelativistic speeds $\beta^{g}_{\rm s}>0.01\,c$, which appear to be superluminal with $\beta_{\rm s}>1$ in the observer frame. In the case of the slowest transverse wave (dashed line), the wave twist angle and transverse velocity of C7 must be three times larger, $\approx 1.5\degr$ and $\approx 0.03\,c$, to measure apparent speeds exceeding the speed of light. We estimate the mean and dispersion of apparent speeds of C7, $\beta_{\rm s} = 9.32 \pm 0.96$ and $\sigma_{\beta_{\rm s}} = 8.06$, for smaller observing intervals less than a month implying that the latter describe more realistic trajectories. Then possible transverse speeds of C7 and twist angles of waves lie in the ranges $(0.08-0.28)\,c$ and $4\degr-14\degr$, respectively (see Fig. \ref{fig:beta_s-phi}). It should be noted that these estimates are obtained for the case when the transverse wave propagates in the sagittal plane, which causes C7 to perform a linear reversing movement along the jet axis. An example close to such a movement is the linear reversal marked in red between 2007.36-2009.15 (Fig.~\ref{fig:MA_dtau=1_clean}). In fact, in most cases we observe nonlinear reversals indicating that the plane of the transverse waves makes some angle with the sagittal plane and/or the transverse waves have a twist in space. Thus, our speed estimates should be taken as evidence that relativistic transverse waves are capable of producing the apparent superluminal speeds of C7.

\paragraph{Intrinsic apparent speed of C7.} The transverse velocity of C7 in the frame of the host galaxy $\beta^{g}_{\rm s, tr}$ is a function of the velocity of a transverse wave, the C7 intrinsic apparent transverse velocity and the viewing angle of the jet (Eq. \ref{eq:beta^g_s}). For estimating the $\beta^{g}_{\rm s, tr}$ we adopt $\theta = 6\degr$ and range of transverse wave speeds discussed above $\beta^{g}_{\rm wave} = 0.979-0.998$. The intrinsic apparent speeds $\beta_{\rm s} = r_{\rm s}/\Delta t$ cannot be directly measured, but its statistical characteristics such as the mean $\overline{\beta}_{\rm s} = \overline{r_{\rm s}} / \overline{\Delta t} $ and standard deviation $\sigma_{\beta_{\rm s}}$ can be estimated. We use the statistical approach developed in \cite{arshakian20} to estimate the unbiased mean and standard deviation of C7 displacements from their Eqs.~(9,10):
\begin{equation}
    \overline{r_{\rm s}} = \frac{\pi}{2} \overline{r_{\rm n}}
    \label{eq:9}
\end{equation}
and
\begin{equation}
    \sigma_{r_{\rm s}} = \sqrt{ 2 \overline{r_{\rm n}^2} - \left(\frac{\pi}{2} \overline{r_{\rm n}}\right)^2 },
    \label{eq:10}
\end{equation}
where $\overline{r_{\rm n}}$ is the mean of the measured displacement projections transverse to the jet central axis and $\overline{r_{\rm n}^2}$ is the mean of the $r_{\rm n}$ squared. They argued that more realistic displacements are those with smaller observation intervals $\Delta t < 35$~days and used them to estimate the $\overline{r_{\rm s}} = 0.015$~mas and $\sigma_{r_{\rm s}} = 0.011$~mas for 54 observational intervals (see their Table~1). Adopting the same constraint, we choose 63 of the 163 observation intervals of our sample and obtain the same values for $\overline{r_{\rm s}}$ and $\sigma_{r_{\rm s}}$ and the mean observation interval $\overline{\Delta t} = 0.051$~yr and $\sigma_{\Delta t} = 0.026$~yr. Then the C7 mean apparent speed $\overline{\beta}_{\rm s} = \overline{r_{\rm s}} / \overline{\Delta t} \approx 1.26$ is superluminal with $\sigma_{\beta_{\rm s}} = 1.13$ using the error propagation in the ratio. For 35 shorter observation intervals limited to $\Delta t < 20$ days, the mean speed is $\overline{\beta}_{\rm s} = 2.5 \pm 0.33$ and $\sigma_{\beta_{\rm s}} = 1.98$. For our analysis, we adopt the latter values as more realistic. 

The mean apparent speed of C7 $\overline{\beta_{\rm s}} = 2.5$ corresponds to a mean speed of $\overline{\beta}^g_{\rm s,tr} \approx 0.05$ and $\overline{\varphi} \approx 2.5\degr$ in the host galaxy frame (Fig.~\ref{fig:beta_s-phi}). Given $\pm 3\sigma_{\overline{\beta}_{\rm s}} \approx \pm 1$ variation in mean speed, we measure ranges of $\overline{\varphi} \approx 0.5\degr - 5.5\degr$ and $\overline{\beta}^g_{\rm s,tr} \approx 0.02-0.11$. The upper limits of the twist angles and speeds of $\varphi \lesssim 12\degr$ and $\beta^g_{\rm s,tr} \lesssim 0.25$ (the latter is still subluminal) are given at $3\sigma_{\beta_{\rm s}} \approx 6$ level. If the apparent speed is subluminal $\beta_{\rm s}<1$ then $\varphi\lesssim 1.5\degr$ and $\beta^g_{\rm s,tr} \lesssim 0.03$. Note that these estimates are model-dependent and may vary with changes in the adopted values of transverse wave velocity and jet viewing angle.

\section{Discussion}
\label{sec:discussions}
The averaging filter and trajectory refinement algorithm smooth the structures on spatial scales of $\lesssim 0.01$~mas. Because of the asymmetric positioning errors of C7, refinement of its trajectory affects structures extended in the jet direction to a greater extent than those in the transverse jet direction. 
For example, the turn enveloping the median center of C7 positions is elongated in the jet direction (upper left panel in Fig.~\ref{fig:MA_dtau=1}), and after trajectory refinement, the reversal turns into an arc-shaped trajectory (upper left panel in Fig.~\ref{fig:MA_dtau=1_clean}). The refinement procedure can affect the reversal type, for example, changing it from RL- to R-type, as in the case of an R-type reversal (red line in the upper right panel of Fig.~\ref{fig:MA_dtau=1_clean}) that looked like an RL-type before refinement (upper right panel of Fig.~\ref{fig:MA_dtau=1_clean}). Trajectory refinement does not affect the direction (CW or CCW) of the reversal. 



The jet ridge line and moving components appear downstream from C7, the median position of which is at a distance of 0.26~mas from the 15~GHz core \citep{cohen14,cohen15}. At high resolution based on a single epoch observation at 86~GHz the jet ridge line is detected up to a projected distance of about 0.4~mas from the core \citep{kim23}. We believe that the component at the core separation of approximately 0.3~mas is the C7 quasi-stationary component. The reasoning behind it is that this component is the brightest feature in the 86~GHz jet beyond 0.1~mas from the core and is situated at the expected core separation taking into account frequency-dependent core shift effect \citep[e.g.,][]{Lobanov98,Sokolovsky11,Pushkarev12}. The jet at 86~GHz has a complex structure with multiple bends within 0.4~mas from the core. This suggests that the transverse waves are generated upstream of C7 and become visible at 15~GHz downstream of C7. The magnetic fields upstream and downstream of C7 have a strong toroidal component, indicating that there is a dense spiral field at sub-parsec \citep{kim23} and pc-scales \citep{pushkarev23}, and that the magnetic field still dominates the jet dynamics at these scales. This argues in favor of an MHD wave model \citep{meier01} in which C7 is the "master" recollimation shock wave, while the moving jet components are compressions of slow and fast magnetosonic waves generated beyond the recollimation shock \citep[][and references therein]{cohen14}, and the transverse patterns of the ridge line are Alfvén MHD waves excited by C7 \citep{cohen15}. The latter is analogous to the excitation of a wave on a whip by shaking the handle. The C7 model proposed in section~\ref{sec:C7 model} assumes that the transverse waves are generated upstream of C7, somewhere between the base of the jet and C7. This diverges from the shaking whip model, where C7 is the site of wave excitation.

Disturbance of the plasma across the jet leads to the excitation of transverse Alfvén waves \citep{gurnett05} that propagate downstream at relativistic speed, similar to hose waves excited under high pressure. In favor of this scenario is the random nature of the twisting wave direction (or reversal direction) discussed earlier. The characteristics of the jet wave, such as amplitude, velocity, and frequency of generation, depend on the characteristics of the toroidal magnetic fields and the jet plasma. 

We assume that during the active state of the jet, powerful transverse waves with large amplitudes of $\lesssim 0.05$~mas are generated, which we observe as irregular C7 reversal patterns in random directions across the jet. When the plasma disturbance weaken, the steady state of the jet comes in the form of quasi-oscillatory waves with amplitude $\lesssim 0.02$~mas during 2010-2012 \citep{cohen15,arshakian20}, which we observe as quasi-oscillatory C7 motion on spatial scales of $\approx 0.01$~mas between 2009.56 and 2011.70 (see the yellow line in Fig.~\ref{fig:MA_dtau=1_clean} and Table~\ref{table:2}). We speculate that the quasi-oscillatory waves observed during the steady state of the jet reflect the wobbling of the inner jet near the accretion disk as seen in 3-dimensional MHD simulations of the black hole-accretion disk-jet system \citep{mckinney13}. Modeling of relativistic MHD waves is necessary to shed light on the physical conditions of the plasma and magnetic field configuration required for the transition from the stable jet state to the active jet state, and to determine the region of wave excitation.

The excited Alfvén waves propagate along the spiral magnetic "muzzle" of the jet and pass through the location of the recollimation shock wave C7. In this scenario, C7 acts as a jet nozzle whose dynamics and direction are controlled by transverse wave characteristics such as amplitude, twist angle, and velocity. The brightness of C7 depends on the intrinsic brightness of the jet beam, its velocity, and the viewing angle of the jet at the C7 position \citep{arshakian20}. The latter varies with the angle of wave twist at C7. Downstream of C7, the jet opening angle is about $4\degr$, and the shape is parabolic, which changes to conical at a distance of 2.5~mas ($\sim 3$~pc) from the core \citep{pushkarev17,kovalev20}. The size of the C7 positional spread ($\sim 0.1$ mas) is determined by the maximum amplitude of the transverse wave ($\sim 0.05$ mas).

The advantage of studying quasi-stationary components in light of the new model is that the C7 characteristics allow us to study the transverse jet wave dynamics at small spatial scales, which are unattainable when analysing smoothed jet ridge lines \citep{cohen14}, as it is constructed in the image domain, while in this work we derive the positions of C7 from the data in the Fourier plane achieving significantly higher effective resolution.

We believe that the study of the dynamics and brightness variations of the QSC can give detailed physical and statistical characterizations of relativistic transverse waves. The QSC dynamics can be strongly influenced by the effect of the core displacements. It is desirable that the dispersion of the core displacements be relatively small compared to the mean intrinsic displacement of the QSC. This allows us to use smaller smoothing intervals and hence to study the dynamics on small temporal and spatial scales. If a QSC is too close to the core, a flux density leakage can easily occur between them, affecting the model fitting results by increasing the corresponding uncertainties. Therefore, to effectively study the dynamics of the QSC, it should be compact and bright, leading to small position errors, and have long-term VLBI observations with periods of significantly high cadence to study the dynamics on different timescales and assessing the true position errors based on results from the nearest observing epochs.

\section{Summary}
We study the intrinsic trajectory of the quasi-stationary component of the jet in BL Lac on sub-parsec scales using 164 epochs of observations made with the VLBA at 15~GHz over 20 years within the MOJAVE and VLBA 2~cm Survey. We highlight the following key results:
\begin{itemize}
    \item To recover the C7 intrinsic trajectory, we apply a moving average procedure to the apparent C7 trajectory, which smooths out the effect of the core displacement caused by the non-stationary nature of the jet. The optimal smoothing interval is about one year.
    
    \item The overall smoothed and refined C7 trajectory is a sequence of reversible trajectories of different spatial scales. We identified 22 reversals, most of which are of R-type. Reversal motion occurs on average every 1.5 years. The number of reversals with clockwise and counterclockwise motion (13 and 9, respectively) suggests that the trajectory twist is random. The distributions of radial distances and azimuthal angles of reversal turn points are not statistically different for clockwise and counterclockwise reversals. We determined that a combination of successive reversals can form a larger RC-type reversal or a quasi-oscillatory trajectory identified as an RO-type reversal.

    \item To describe the reversible motion of C7, we propose a model in which C7 is a standing recollimation shock wave and its dynamics is governed by the passage of relativistic transverse waves (Alfvén waves) through the stationary component of C7, which moves in a plane normal to the central axis of the jet, similar to the motion of a seagull on a wave. In the active state of the jet, irregular transverse waves with random twist are excited. The jet stabilizes into quasi-oscillatory waves with regular twist, which is observed as a quasi-oscillatory motion of C7. We estimate the mean wave frequency to be about 1.5 per year, which is the upper limit. The amplitude of the transverse waves ranges approximately from 0.01 to 0.05~mas ($0.013-0.065$~pc).

    \item 
    The measured mean superluminal speeds of C7 can be understood within the framework of the proposed model. The relativistic transverse wave moves C7 at subluminal speed ($\sim~0.04\,c$) in the host galaxy's reference frame, while the observer detects it as superluminal $\sim 2\,c$. The apparent transverse speed of C7 is superluminal for a maximum wave speed of $0.998\,c$ (or Lorentz factor $\Gamma_{\rm wave} = 15$) and wave twist angle at the C7 position of $\gtrsim 0.5\degr$.

    \item 
    Important implications of the C7 model are that relativistic transverse waves are generated upstream of QSC and are mostly twisted in space, and the dynamics and kinematics of the QSC are determined by the speed, amplitude, and twist angle of the transverse wave, the maximum wave amplitude is about half of the QSC scattering size, and the superluminal speeds of the QSC indicate the presence of relativistic transverse waves in the jet. 
    
\end{itemize}

\begin{acknowledgements}
We thank Yuri Y. Kovalev, Andrei Lobanov and Eduardo Ros for valuable comments and discussions. The VLBA is a facility of the National Radio Astronomy Observatory, a facility of the National Science Foundation that is operated under cooperative agreement with Associated Universities, Inc. This research has made use of data from the MOJAVE database that is maintained by the MOJAVE team \citep{lister18}. The research was supported by the YSU, in the frames of the internal grant. A.B.P. is supported in the framework of the State project ``Science'' by the Ministry of Science and Higher Education of the Russian Federation under the contract 075-15-2024-541.
\end{acknowledgements}





\bibliography{refs_tga}
\bibliographystyle{aa} 

\end{document}